\documentclass{article}
\usepackage{spconf}
\usepackage{amsmath, amsfonts, graphicx, booktabs, tabularx, multirow, siunitx, xcolor, soul, bm, enumitem, tikz, mathtools, subcaption}
\usepackage{hyperref}

\usepackage{enumitem}
\setlist{nosep, leftmargin=14pt}

\graphicspath{{./figures/}}

\title{FNOSeg3D: Resolution-Robust 3D Image Segmentation with\\ Fourier Neural Operator}
\name{Ken C. L. Wong$^{*}$, Hongzhi Wang, Tanveer Syeda-Mahmood
\thanks{This paper was accepted by IEEE ISBI 2023. \copyright 2023 IEEE. Personal use of this material is permitted. Permission from IEEE must be obtained for all other uses, in any current or future media, including reprinting/republishing this material for advertising or promotional purposes, creating new collective works, for resale or redistribution to servers or lists, or reuse of any copyrighted component of this work in other works.
}
\thanks{
\noindent$^{*}$Corresponding author (clwong@us.ibm.com)
}
}
\address{IBM Research -- Almaden Research Center, San Jose, CA, USA}
\begin{document}
%
\maketitle
\begin{abstract}
Due to the computational complexity of 3D medical image segmentation, training with downsampled images is a common remedy for out-of-memory errors in deep learning. Nevertheless, as standard spatial convolution is sensitive to variations in image resolution, the accuracy of a convolutional neural network trained with downsampled images can be suboptimal when applied on the original resolution. To address this limitation, we introduce FNOSeg3D, a 3D segmentation model robust to training image resolution based on the Fourier neural operator (FNO). The FNO is a deep learning framework for learning mappings between functions in partial differential equations, which has the appealing properties of zero-shot super-resolution and global receptive field. We improve the FNO by reducing its parameter requirement and enhancing its learning capability through residual connections and deep supervision, and these result in our FNOSeg3D model which is parameter efficient and resolution robust. When tested on the BraTS'19 dataset, it achieved superior robustness to training image resolution than other tested models with less than 1\% of their model parameters\footnote{The GitHub repository is available at \url{https://github.com/IBM/multimodal-3d-image-segmentation}}.
\end{abstract}
\section{Introduction}

With the introduction of convolutional neural networks (CNNs), the accuracy and speed of medical image segmentation have been significantly improved \cite{Journal:Hesamian:DI2019:deep,Journal:Liu:Sustainability2021:review}. Nevertheless, given the computationally intensive nature of CNNs, out-of-memory errors in GPU are usually encountered in 3D image segmentation. Comparing to 2D segmentation, segmentation in 3D is more resource demanding as the number of feature components in each layer can be orders of magnitude larger. To remedy the out-of-memory errors, image downsampling and patch-wise training are common approaches to reduce the input image size. As spatial convolution is sensitive to variations in image resolution, the accuracy of CNNs trained with downsampled images can be suboptimal when applied on the original image resolution. On the other hand, although patch-wise training can maintain the original image resolution, the receptive field can be largely reduced depending on the patch size. Furthermore, post-processing is usually required to combine the patch-wise predictions together. Therefore, there are different tradeoffs with different approaches.

Regardless of the approaches for reducing input sizes, the computational complexity still limits the learning capability of long-range spatial dependencies of 3D CNNs. As the memory usage is proportional to the width and depth of the model, the numbers of filters and layers can be used are much smaller in 3D segmentation. Apart from constraining the level of abstraction, the limitation on the number of layers also constrains the overall receptive field given the local receptive fields of convolutional layers. To address this issue, transformers are introduced to medical image segmentation \cite{Conference:Gao:MICCAI2021:utnet,Conference:Xie:MICCAI2021:cotr,Conference:Hatamizadeh:WACV2022:unetr}. These approaches form a sequence of samples by either dividing an image into smaller patches or by using the pixel values of low-resolution features, and the sequence is input to the multi-head attention module of transformers to learn the long-range spatial dependencies. Although the results are promising, as the computational requirements of transformers are proportional to sequence lengths which increase with image sizes, size-reduction approaches are still required.

In view of these issues, a model that is robust to training image resolution and has a global receptive field is desirable. Towards this direction, here we propose a 3D segmentation model, FNOSeg3D, that can provide accurate results when applied on images with higher resolutions than the training images (i.e., zero-shot super-resolution). This model is based on the Fourier neural operator (FNO) which is a deep learning model that learns mappings between functions in partial differential equations (PDEs) \cite{Conference:Li:ICLR2021:fourier}. As the formulations of FNO were derived based on the idea of the Green's function in continuous space, it possesses the appealing properties of zero-shot super-resolution and global receptive field. Our contributions in this paper include:
\begin{itemize}
  \item We modify FNO for computationally expensive 3D medical image segmentation. We reduce the parameter requirement of FNO and enhance its segmentation performance. These result in a resolution-robust 3D segmentation model that has orders of magnitude fewer parameters than most deep learning 3D segmentation models.
  \item We compare our model with other models on different training image resolutions to study their robustness. This provides useful insights that are usually unavailable in other studies.
\end{itemize}
We tested our framework on the Multimodal Brain Tumor Segmentation Challenge 2019 (BraTS'19) dataset \cite{Journal:Bakas:arXiv2018:identifying}. When trained with the original image resolution, FNOSeg3D achieved an average Dice coefficient of 79\%, which is similar to other tested models but only had 29.8k parameters, less than 1\% of other tested models. FNOSeg3D also showed superior robustness on training image resolution.

\section{Methodology}

\subsection{Fourier Neural Operator}

Our model is based on the Fourier neural operator, which is a deep learning model proposed to learn mappings between functions in PDEs without knowing the actual PDEs \cite{Conference:Li:ICLR2021:fourier}. As its formulations were developed based on the Green's function in the continuous space, the model can learn a single set of network parameters to be used with different resolutions. Such zero-shot super-resolution capability is desirable for computationally demanding 3D image segmentation as a model trained with lower-resolution images can be applied on higher-resolution images with decent accuracy. The neural operator is formulated as iterative updates:
\begin{equation}\label{eq:fno_update}
    \begin{gathered}
      v_{t+1}(x) \coloneqq \sigma \left(W v_t(x) + \left(\mathcal{K}v_t\right)(x)\right) \\
      \textrm{with} \ \ \left(\mathcal{K}v_t\right)(x) \coloneqq \int_{D} \kappa(x - y)v_t(y) \,dy, \ \ \forall x \in D
    \end{gathered}
\end{equation}
where $v_t(x) \in \mathbb{R}^{d_{v_t}}$ is a function of $x$. $W \in \mathbb{R}^{d_{v_{t+1}} \times d_{v_t}}$ is a learnable linear transformation and $\sigma$ accounts for normalization and activation. In our work, $D \subset \mathbb{R}^3$ represents the 3D imaging space, and $v_t(x)$ are the outputs of hidden layers with $d_{v_t}$ channels. $\mathcal{K}$ is the kernel integral operator with $\kappa \in \mathbb{R}^{d_{v_{t+1}} \times d_{v_t}}$ a learnable kernel function. As $\left(\mathcal{K}v_t\right)(x)$ is a convolution operator, its efficiency can be improved by applying the convolution theorem, which states that the Fourier transform ($\mathcal{F}$) of a convolution of two functions is the pointwise product of their Fourier transforms:
\begin{equation}\label{eq:fourier_conv}
    \begin{split}
      \left(\mathcal{K}v_t\right)(x) &= \mathcal{F}^{-1}\left(\mathcal{F}(\kappa) \cdot \mathcal{F}(v_t)\right)(x) \\
      &= \mathcal{F}^{-1}\left(R \cdot \left(\mathcal{F}v_t\right)\right)(x), \ \ \forall x \in D
    \end{split}
\end{equation}
$R(k) \in \mathbb{C}^{d_{v_{t+1}} \times d_{v_t}}$ is a learnable function in the Fourier domain and $\left(\mathcal{F}v_t\right)(k) \in \mathbb{C}^{d_{v_t}}$. As the fast Fourier transform is used in implementation, $k \in \mathbb{N}^3$ are non-negative integer coordinates, and each $k$ has a learnable $R(k)$. Only $k_i \leq k_{\mathrm{max},i}$ corresponding to the lower frequencies in each dimension $i$ are used to reduce model parameters and computation time.

\begin{figure}[t]
    \centering
    \includegraphics[width=\linewidth]{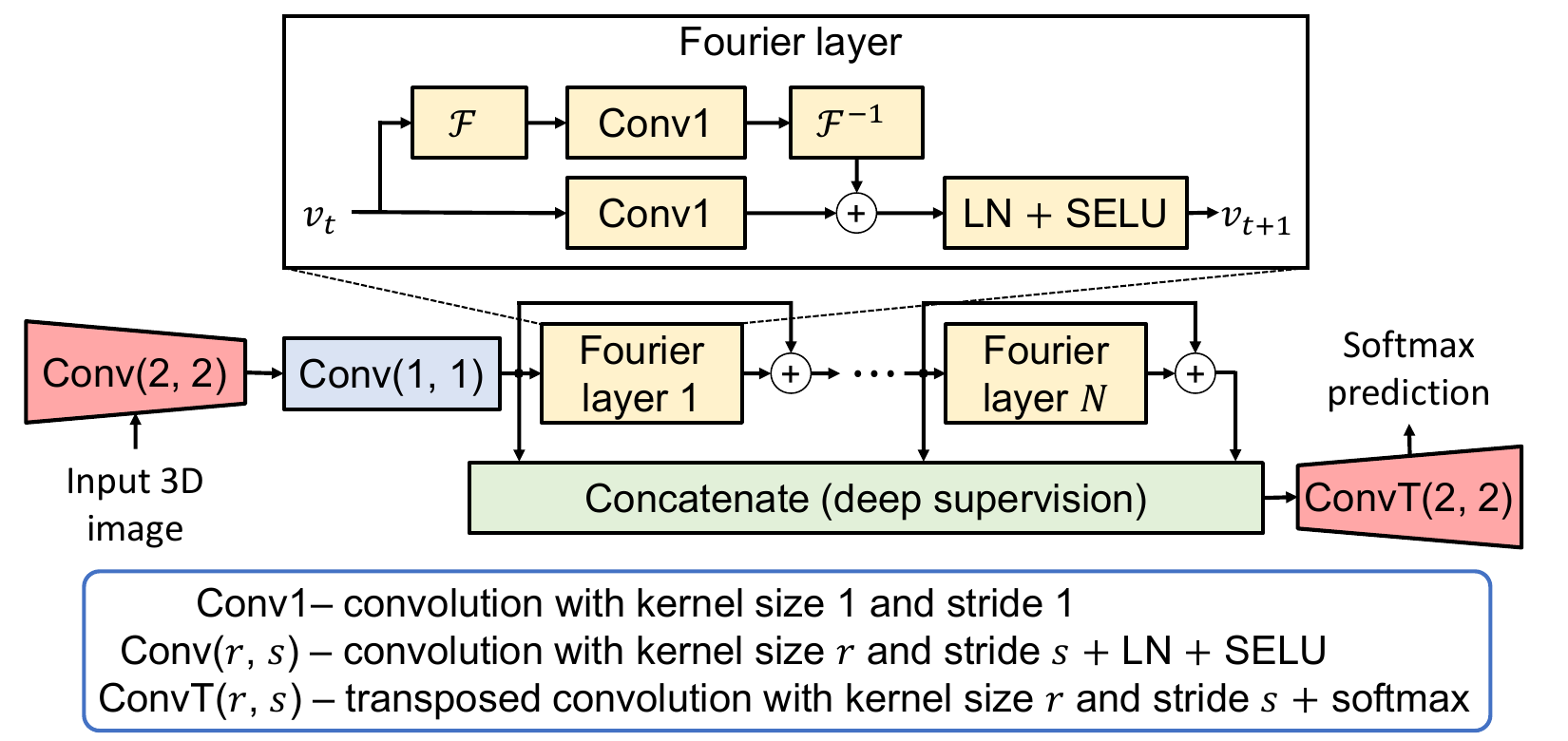}
    \caption{Network architecture of FNOSeg3D. The Fourier layers are (\ref{eq:fno_update}) implemented by the fast Fourier transform, with $d_{v_{t+1}}$ = $d_{v_{t}}$ = 12, $k_\mathrm{max}$ = (15, 15, 10) in the ($x$, $y$, $z$) directions, and $N$ = 32 in our experiments. The red blocks are for learnable resampling. The original FNO is FNOSeg3D without deep supervision and residual connections, and the weights are not shared in the Fourier domain.}
    \label{fig:network}
\end{figure}

\subsection{FNOSeg3D -- Parameter-Efficient FNO for Resolution-Robust 3D Image Segmentation}
\label{sec:network}

Here we propose FNOSeg3D by modifying the original FNO for 3D image segmentation (Fig. \ref{fig:network}). The original FNO uses a different $R(k)$ for each $k$ in (\ref{eq:fourier_conv}), and this results in a large number of model parameters in complex numbers even when $k_\mathrm{max}$ is much smaller than the image size. To remedy this, we use the same (shared) $R(k)$ for all $k$ to significantly reduce the number of parameters and improve the accuracy suffered from over-parameterization. This is equivalent to 3D convolution with a kernel size of one in the Fourier domain with complex numbers. Furthermore, residual connections \cite{Conference:He:CVPR2016} and deep supervision \cite{Conference:Lee:AISTATS2015} are used to improve the training stability, convergence, and accuracy. As the batch size is usually small because of the large memory requirement for 3D segmentation, layer normalization (LN) is used \cite{Journal:Ba:arXiv2016:layer}. The scaled exponential linear unit (SELU) \cite{Conference:Klambauer:NIPS2017:self} is used as the activation function, and the softmax function is used to produce the final prediction scores. The spatial integration in (\ref{eq:fno_update}) realized by the Fourier transform provides a global receptive field as all voxels are used to compute the value at each $k$. Thus, pooling is not required in the network architecture.

As using the original image resolution usually results in out-of-memory errors in 3D segmentation, downsampling the inputs and then upsampling the predictions are usually required. Instead of using traditional image resampling methods, we use a convolutional layer with the kernel size and stride of two right after the input layer, and replace the final output convolutional layer by a transposed convolutional layer with the kernel size and stride of two (red blocks in Fig. \ref{fig:network}). In this way, the model can learn the optimal resampling approach. All these modifications result in the simple yet powerful architecture of FNOSeg3D.

\subsection{Loss Function with Pearson's Correlation Coefficient (PCC)}

The PCC loss ($L_{PCC} \in [0, 1]$) is used as it is robust to learning rate and accurate for image segmentation \cite{Workshop:Wong:MLMI2022:3d}, and it consistently outperformed the Dice loss and weighted cross-entropy in our experiments. The $L_{PCC}$ is computed as:
\begin{gather}
\label{eq:pcc_loss_1}
L_{PCC} = \mathbf{E}[1 - PCC_l], \\
\label{eq:pcc_loss_2}
PCC_l = 0.5 \left(\tfrac{\sum_{i=1}^{N}(p_{li} - \bar{p}_l)(y_{li} - \bar{y}_l)}{\sqrt{\left(\sum_{i=1}^{N}(p_{li} - \bar{p}_l)^2\right)\left(\sum_{i=1}^{N}(y_{li} - \bar{y}_l)^2\right)  + \epsilon}} + 1 \right)
\end{gather}
with $\mathbf{E}[\bullet]$ the mean value across semantic labels $l$. $p_{li} \in [0, 1]$ are the prediction scores, $y_{li} \in \{0, 1\}$ are the ground-truth values, and $N$ is the number of voxels in an image. $\epsilon$ is a small positive number (e.g., $10^{-7}$) to avoid the divide-by-zero situations, for example, when label $l$ is missing in an image and all $y_{li} = 0$. $L_{PCC}$ = 0, 0.5, and 1 represent perfect prediction, random prediction, and total disagreement, respectively. As the means are subtracted from the samples in (\ref{eq:pcc_loss_2}), both scores of the foreground and background voxels of each label contribute to $L_{PCC}$. Hence, a low $L_{PCC}$ is achievable only if both foreground and background are well classified.

\begin{figure}[t]
    \centering
    \includegraphics[width=0.6\linewidth]{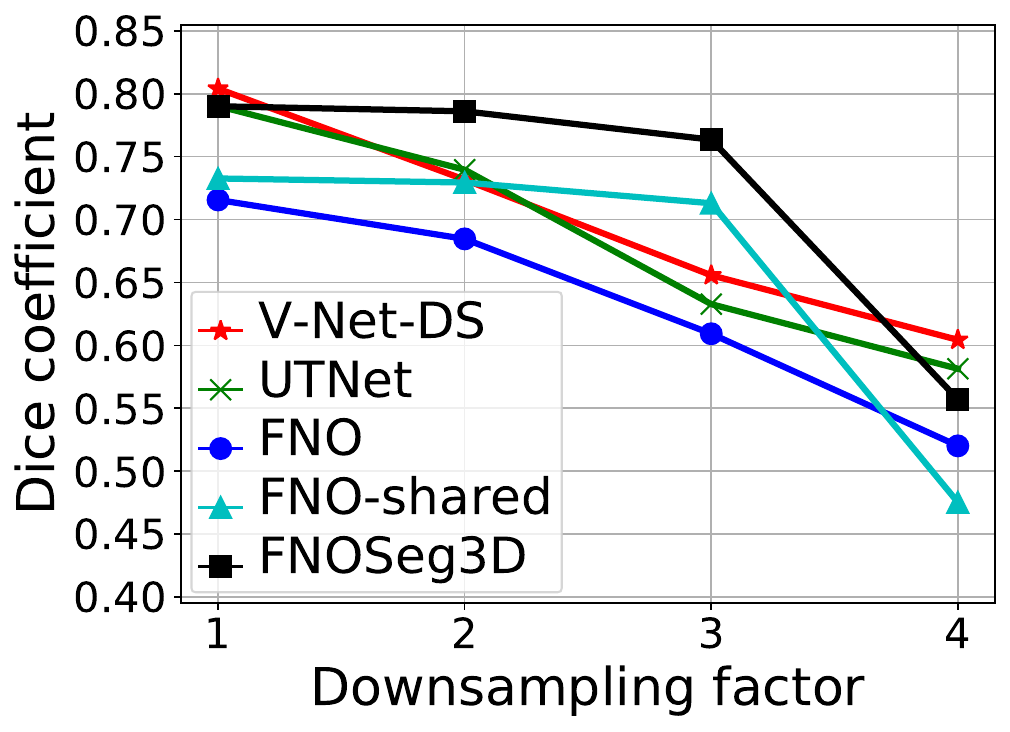}
    \caption{Comparisons of robustness to training image resolution. Each point represents the average value from WT, TC, and ET of the 125 official validation cases of BraTS'19. The training images were downsampled by different factors while the trained models were tested with the original resolution.}
    \label{fig:vs_downfactor}
\end{figure}

\subsection{Training Strategy}

For multimodal data, the images of different modalities are stacked along the channel axis to provide a multi-channel input. Image normalization is performed on each image of each modality as the intensity ranges across modalities can be quite different. Image augmentation with rotation (axial, $\pm$\ang{30}), shifting ($\pm$20\%), and scaling ([0.8, 1.2]) is used and each image has an 80\% chance to be transformed. The Adamax optimizer \cite{Journal:Kingma:arXiv2014} was used with the cosine annealing learning rate scheduler \cite{Conference:Loshchilov:ICLR2017:SGDR}, with the maximum and minimum learning rates as $10^{-2}$ and $10^{-3}$, respectively. An NVIDIA Tesla P100 GPU with 16 GB memory is used with a batch size of one and 100 epochs. Note that small batch sizes are common in 3D segmentation given the large memory requirement.

\section{Experiments}
\label{sec:experiments}

\subsection{Data and Experimental Setups}

The dataset of BraTS'19 was used \cite{Journal:Bakas:arXiv2018:identifying}. The training dataset has 335 cases of gliomas, each with four modalities of T1, post-contrast T1, T2, and T2-FLAIR images of size 240$\times$240$\times$155. The non-overlapping annotations comprise ``enhancing tumor'', ``peritumoral edema'', and ``necrotic and non-enhancing tumor core''. There is also an official validation dataset of 125 cases in the same format without given annotations.

To study the robustness to image resolution, models were trained with images downsampled by different factors (1, 2, 3, and 4). In training, we split the training dataset (335 cases) into 90\% for training and 10\% for validation. In testing, regardless of the downsampling factor, each model was tested on the official validation dataset (125 cases) with 240$\times$240$\times$155 voxels. The predictions were uploaded to https://ipp.cbica.upenn.edu/ for the Dice coefficients of the ``whole tumor'' (WT), ``tumor core'' (TC), and ``enhancing tumor'' (ET) regions which are formed by combining the non-overlapping annotations \cite{Journal:Bakas:arXiv2018:identifying}. We compare our proposed FNOSeg3D with four other models:
\begin{enumerate}
  \item \textbf{V-Net-DS} \cite{Conference:Wong:MICCAI2018}: a V-Net with deep supervision representing the commonly used architectures.
  \item \textbf{UTNet} \cite{Conference:Gao:MICCAI2021:utnet}: a U-Net with transformer's attention modules representing the recent self-attention architectures.
  \item \textbf{FNO} \cite{Conference:Li:ICLR2021:fourier}: the original FNO.
  \item \textbf{FNO-shared}: FNO with shared weights.
\end{enumerate}
The learnable resampling approach in Section \ref{sec:network} was applied to all models. Note that our goal is not competing for the best accuracy but studying the robustness to image resolution.

\begin{table}[t]
\caption{Numerical comparisons of Dice coefficients (\%) with different training image resolutions. The numbers of parameters are shown next to the model names.}
\label{table:results}

\fontsize{8}{9}\selectfont
\centering

\newcolumntype{C}{>{\centering\arraybackslash}X}
\newcommand{\boldblue}[1]{\textcolor{blue}{\textbf{#1}}}

\begin{tabularx}{\linewidth}{@{\extracolsep{2pt}}lCCCCCC}
\toprule
Downsampling factor & \multicolumn{3}{c}{1 (240$\times$240$\times$155)} & \multicolumn{3}{c}{3 (80$\times$80$\times$52)} \\
\cline{1-1} \cline{2-4} \cline{5-7} \noalign{\smallskip}
Region & \multicolumn{1}{c}{WT} & \multicolumn{1}{c}{TC} & \multicolumn{1}{c}{ET} & \multicolumn{1}{c}{WT} & \multicolumn{1}{c}{TC} & \multicolumn{1}{c}{ET} \\
\midrule
V-Net-DS (5.7M) & 88.8 & 77.7 & 74.7 & 70.6 & 57.9 & 68.2 \\
UTNet (7.1M) & 86.9 & 76.1 & 74.0 & 69.9 & 56.8 & 63.1 \\
FNO (165.9M) & 83.8 & 69.0 & 61.8 & 78.2 & 58.8 & 45.8 \\
FNO-shared (17.2k) & 84.7 & 66.4 & 68.7 & 82.9 & 71.1 & 59.8 \\
FNOSeg3D (29.8k) & 87.8 & 75.8 & 73.4 & 86.4 & 72.3 & 70.4 \\
\bottomrule
\end{tabularx}
\end{table}

\begin{table}[t]
\caption{Inference time per image in second averaged from images of size 240$\times$240$\times$155.}
\label{table:inference_time}

\fontsize{7}{8}\selectfont
\centering

\newcolumntype{C}{>{\centering\arraybackslash}X}
\newcommand{\boldblue}[1]{\textcolor{blue}{\textbf{#1}}}

\begin{tabularx}{\linewidth}{CCCCC}
\toprule
V-Net-DS & UTNet & FNO & FNO-shared & FNOSeg3D \\
\midrule
0.33 & 0.41 & 0.50 & 0.61 & 0.66 \\
\bottomrule
\end{tabularx}
\end{table}

\begin{figure}[t]
\fontsize{6}{7}\selectfont
    \centering
    \begin{minipage}[t]{0.17\linewidth}
      \vspace{24.5em}
      \centering{Ground truth} \\
      \includegraphics[width=\linewidth]{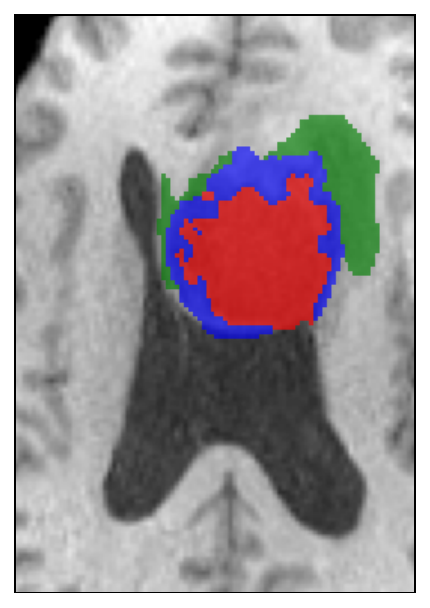}\\
    \end{minipage}
    \begin{minipage}[t]{0.7\linewidth}
        \hrule
        \vspace{0.5em}
        \begin{minipage}[t]{\linewidth}
          \centering{V-Net-DS}
        \end{minipage}
        \begin{minipage}[t]{0.24\linewidth}
          \includegraphics[width=\linewidth]{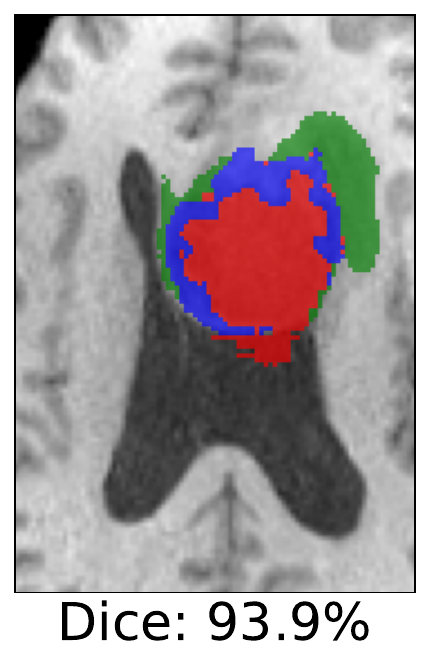}
        \end{minipage}
        \begin{minipage}[t]{0.24\linewidth}
          \includegraphics[width=\linewidth]{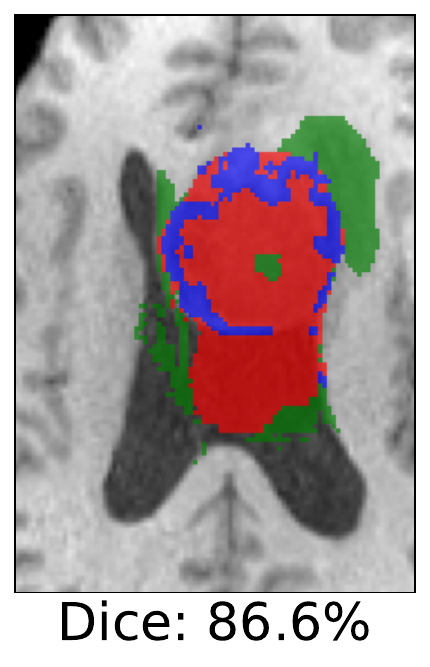}
        \end{minipage}
        \begin{minipage}[t]{0.24\linewidth}
          \includegraphics[width=\linewidth]{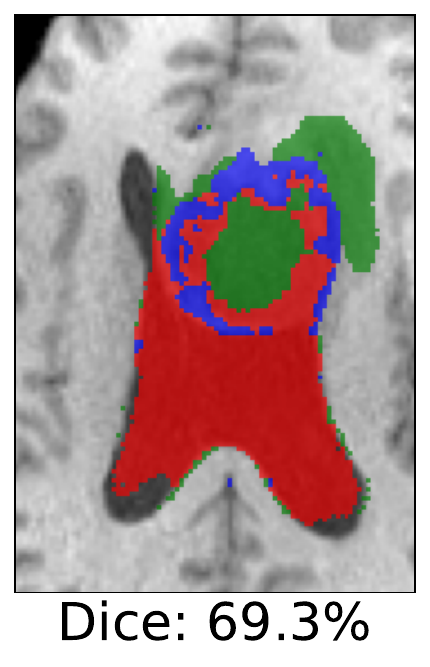}
        \end{minipage}
        \begin{minipage}[t]{0.24\linewidth}
          \includegraphics[width=\linewidth]{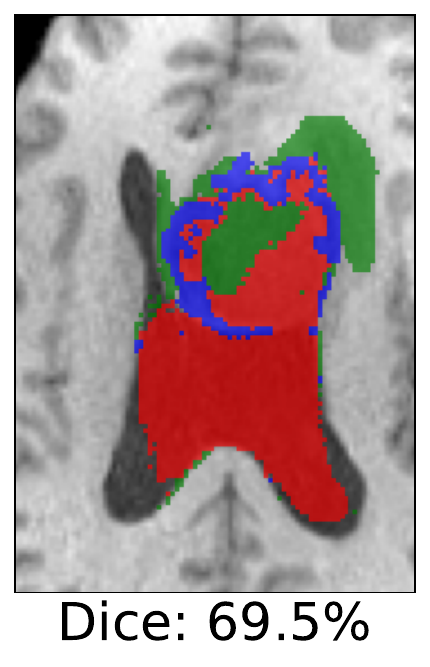}
        \end{minipage}
        \hrule
        \vspace{0.5em}
        \begin{minipage}[t]{\linewidth}
          \centering{UTNet}
        \end{minipage}
        \begin{minipage}[t]{0.24\linewidth}
          \includegraphics[width=\linewidth]{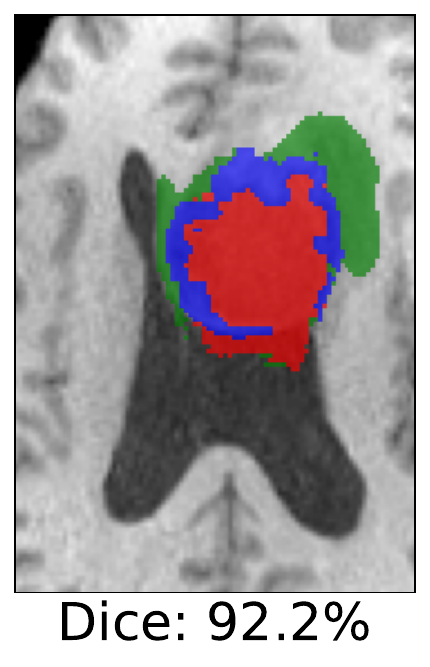}
        \end{minipage}
        \begin{minipage}[t]{0.24\linewidth}
          \includegraphics[width=\linewidth]{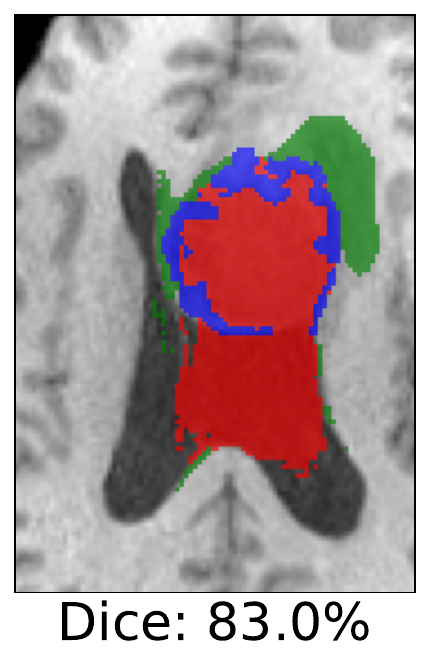}
        \end{minipage}
        \begin{minipage}[t]{0.24\linewidth}
          \includegraphics[width=\linewidth]{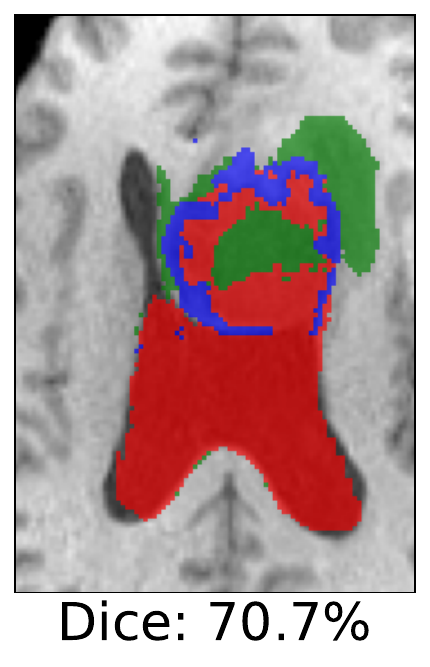}
        \end{minipage}
        \begin{minipage}[t]{0.24\linewidth}
          \includegraphics[width=\linewidth]{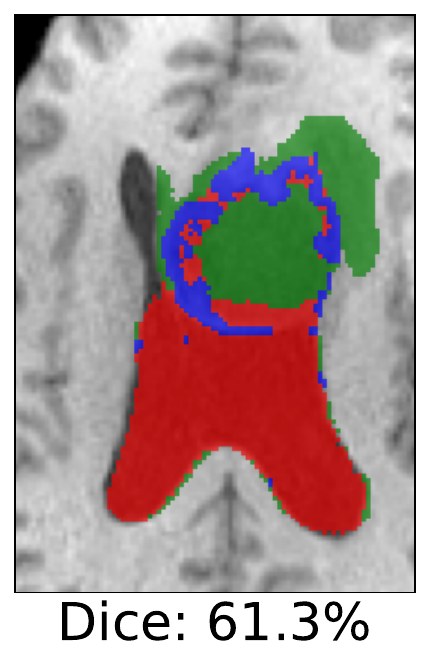}
        \end{minipage}
        \hrule
        \vspace{0.5em}
        \begin{minipage}[t]{\linewidth}
          \centering{FNO}
        \end{minipage}
        \begin{minipage}[t]{0.24\linewidth}
          \includegraphics[width=\linewidth]{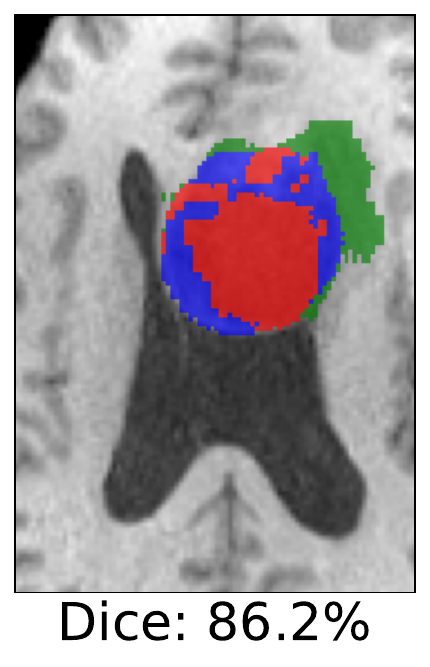}
        \end{minipage}
        \begin{minipage}[t]{0.24\linewidth}
          \includegraphics[width=\linewidth]{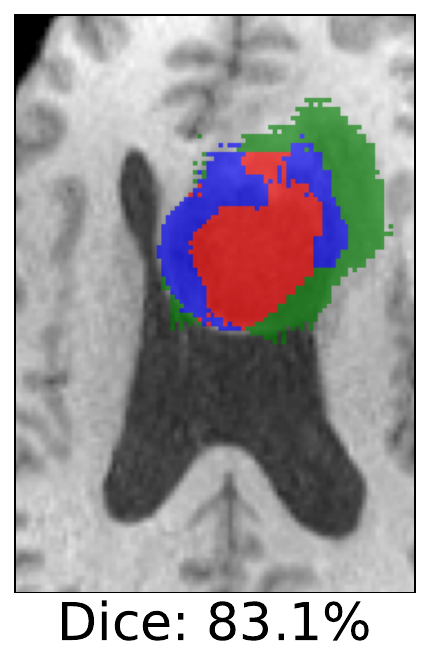}
        \end{minipage}
        \begin{minipage}[t]{0.24\linewidth}
          \includegraphics[width=\linewidth]{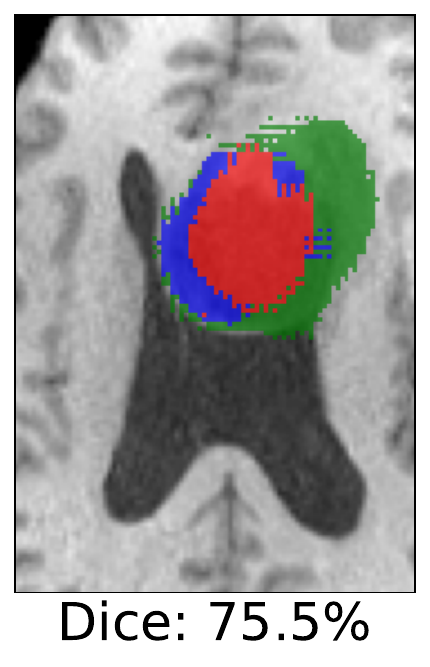}
        \end{minipage}
        \begin{minipage}[t]{0.24\linewidth}
          \includegraphics[width=\linewidth]{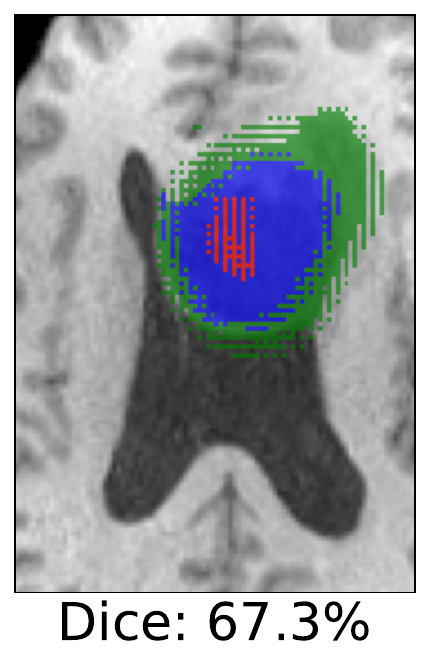}
        \end{minipage}
        \hrule
        \vspace{0.5em}
        \begin{minipage}[t]{\linewidth}
          \centering{FNO-shared}
        \end{minipage}
        \begin{minipage}[t]{0.24\linewidth}
          \includegraphics[width=\linewidth]{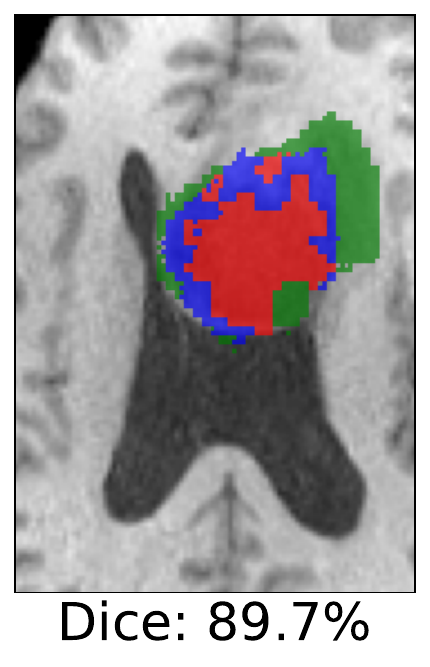}
        \end{minipage}
        \begin{minipage}[t]{0.24\linewidth}
          \includegraphics[width=\linewidth]{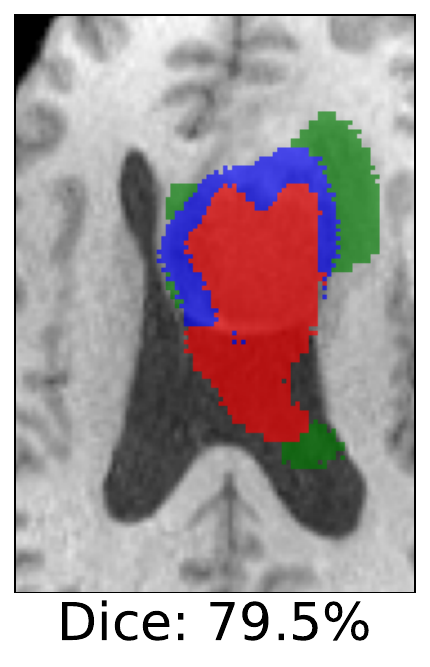}
        \end{minipage}
        \begin{minipage}[t]{0.24\linewidth}
          \includegraphics[width=\linewidth]{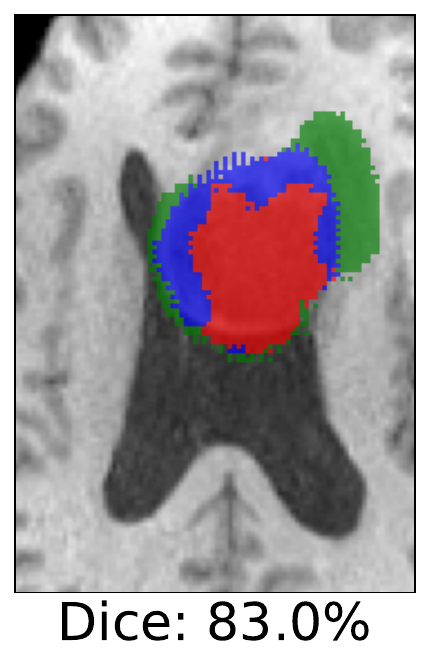}
        \end{minipage}
        \begin{minipage}[t]{0.24\linewidth}
          \includegraphics[width=\linewidth]{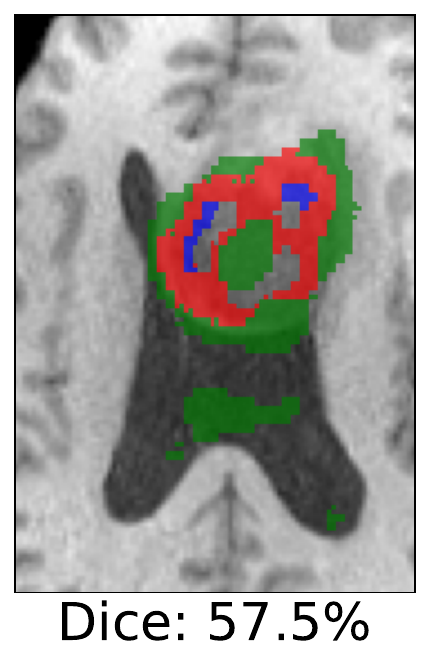}
        \end{minipage}
        \hrule
        \vspace{0.5em}
        \begin{minipage}[t]{\linewidth}
          \centering{FNOSeg3D}
        \end{minipage}
        \begin{minipage}[t]{0.24\linewidth}
          \includegraphics[width=\linewidth]{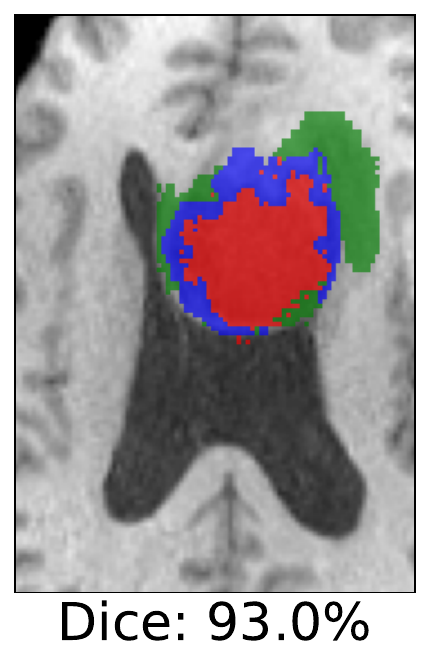}
        \end{minipage}
        \begin{minipage}[t]{0.24\linewidth}
          \includegraphics[width=\linewidth]{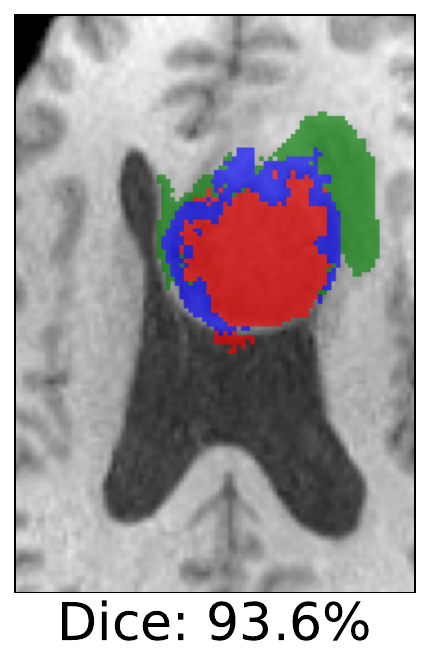}
        \end{minipage}
        \begin{minipage}[t]{0.24\linewidth}
          \includegraphics[width=\linewidth]{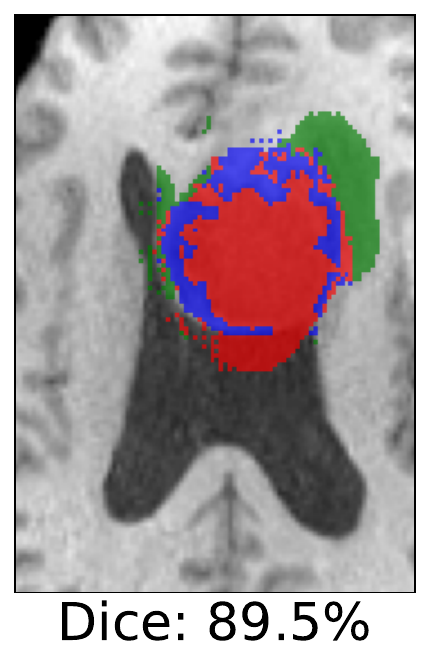}
        \end{minipage}
        \begin{minipage}[t]{0.24\linewidth}
          \includegraphics[width=\linewidth]{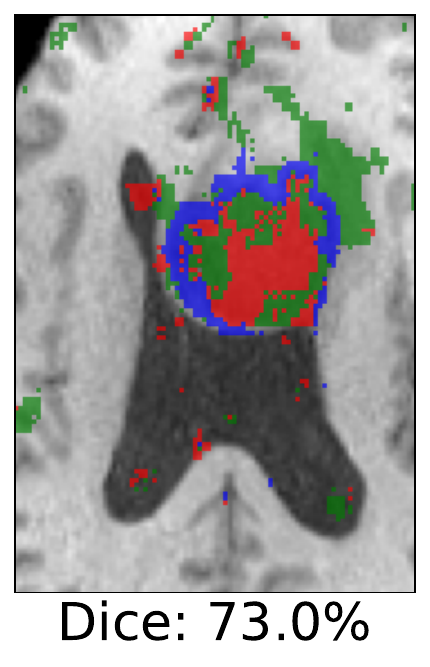}
        \end{minipage}
        \hrule
        \vspace{0.5em}
        \begin{minipage}[t]{0.24\linewidth}
          \centering{240$\times$240$\times$155}
        \end{minipage}
        \begin{minipage}[t]{0.24\linewidth}
          \centering{120$\times$120$\times$78}
        \end{minipage}
        \begin{minipage}[t]{0.24\linewidth}
          \centering{80$\times$80$\times$52}
        \end{minipage}
        \begin{minipage}[t]{0.24\linewidth}
          \centering{60$\times$60$\times$39}
        \end{minipage}
    \end{minipage}
    \caption{Visual comparisons among models trained with different image resolutions, tested on an unseen sample of size 240$\times$240$\times$155. The Dice coefficients were averaged from the WT, TC, and ET regions.}
    \label{fig:visualization}
\end{figure}

\subsection{Results and Discussion}

Table \ref{table:results} and Fig. \ref{fig:vs_downfactor} show the results on the official validation set. Without using shared $R(k)$ in (\ref{eq:fourier_conv}), FNO had 165.9 million parameters, which were over 5,000 times more than FNOSeg3D (29.8 thousand) and had the worst performance. On the other hand, FNO-shared outperformed FNO by simply using shared weights, thus FNO is over-parameterized. FNOSeg3D outperformed FNO-shared with the use of residual connections and deep supervision, but its number of parameters was larger (29.8k vs. 17.2k). When trained with images of 240$\times$240$\times$155 voxels, V-Net-DS outperformed FNOSeg3D by 1.4\% on average but the number of parameters of V-Net-DS (5.7M) was over 190 times larger. UTNet and FNOSeg3D had almost the same accuracy on average but UTNet (7.1M parameters) had over 230 times more parameters. Table \ref{table:inference_time} shows the average inference times of different models. Although FNOSeg3D had the longest inference time, it only took 0.66 second per image of size 240$\times$240$\times$155. Interestingly, the time difference between FNO and FNO-shared shows that some implementation details may be optimized to reduce the inference time of FNOSeg3D.

For the robustness to training image resolution, the accuracy reductions of V-Net-DS and UTNet were almost linear with respect to the downsampling factor (Fig. \ref{fig:vs_downfactor}). In contrast, FNO-shared and FNOSeg3D were more robust and their trends were similar. Table \ref{table:results} shows that when the image size reduced from 240$\times$240$\times$155 to 80$\times$80$\times$52 (downsampling factor = 3), the reductions in Dice coefficients of FNOSeg3D and FNO-shared were less than 3\% on average, while the reductions of V-Net-DS and UTNet were more than 14\%. When the image size reduced to 60$\times$60$\times$39 (downsampling factor = 4), all models failed to learn properly.

Fig. \ref{fig:visualization} shows the visual comparisons among the models trained with different image resolutions and tested on an unseen sample of size 240$\times$240$\times$155. Consistent with Table \ref{table:results} and Fig. \ref{fig:vs_downfactor}, when trained with the original image size, V-Net-DS, UTNet, and FNOSeg3D performed similarly. Nevertheless, the accuracies of V-Net-DS and UTNet quickly decreased with the reduction in the training image size. In contrast, FNOSeg3D was more robust to the training resolution.

Therefore, by inheriting the zero-shot super-resolution and global receptive field properties of FNO, FNOSeg3D trained with lower-resolution images can be applied to higher-resolution images with small reductions in accuracy. Furthermore, with the enhancement of shared weights, residual connections, and deep supervision, decent accuracy can be achieved with surprisingly small numbers of model parameters. These advantages are desirable to the computationally intensive 3D image segmentation. Given its relatively simple network architecture, FNOSeg3D can be combined with other deep learning techniques for further improvements.

\section{Conclusion}

We introduce FNOSeg3D for resolution-robust 3D image segmentation. We improve FNO by reducing the parameter requirement, and enhance its learning capability by residual connections, deep supervision, and the PCC loss. Together with the properties of zero-shot super-resolution and global receptive field of FNO, FNOSeg3D is parameter efficient and robust to training image resolution. Experimental results show that FNOSeg3D can achieve similar performance as V-Net and transformer-enhanced segmentation models when trained with high-resolution images, and can have superior performance when trained with low-resolution images.

\vspace{1em}
\noindent\textbf{Compliance with Ethical Standards:} This research study was conducted retrospectively using human subject data made available in open access by the Multimodal Brain Tumor Segmentation Challenge 2019 (BraTS'19).

\vspace{1em}
\noindent\textbf{Conflicts of Interest:} The authors have no relevant financial or non-financial interests to disclose.

\bibliographystyle{IEEEbib}
\bibliography{Ref}

\end{document}